\documentclass[aps,pra,floatfix,showpacs,twocolumn,nofootinbib,
superscriptaddress]{revtex4}

\usepackage{amsmath}
\usepackage{amssymb}
\usepackage{graphicx}
\usepackage{color}

\newcommand{\ket}[1]{\mbox{$|#1\rangle$}}

\newcommand{\ketbra}[2]{\mbox{$|#1\rangle\langle #2|$}}

\begin{document}

\title{Distance measures to compare real and ideal quantum processes}

\author{Alexei~Gilchrist} \email{alexei@physics.uq.edu.au}
\affiliation{Centre for Quantum Computer Technology and Department of
  Physics, The University of Queensland, Brisbane, Queensland 4072,
  Australia.}

\author{Nathan~K.~Langford} \email{langford@physics.uq.edu.au}
\affiliation{Centre for Quantum Computer Technology and Department of
  Physics, The University of Queensland, Brisbane, Queensland 4072,
  Australia.}

\author{Michael~A.~Nielsen} \email{nielsen@physics.uq.edu.au}
\homepage{www.qinfo.org/people/nielsen} \affiliation{School of
  Physical Sciences and School of Information Technology and
  Electrical Engineering, The University of Queensland, Brisbane,
  Queensland 4072, Australia.}

\date{\today}

\begin{abstract}
  With growing success in experimental implementations it is critical
  to identify a ``gold standard'' for quantum information processing,
  a single measure of distance that can be used to compare and
  contrast different experiments.  We enumerate a set of criteria such
  a distance measure must satisfy to be both experimentally and
  theoretically meaningful.  We then assess a wide range of possible
  measures against these criteria, before making a recommendation as
  to the best measures to use in characterizing quantum information
  processing.
\end{abstract}


\maketitle
        
%
%

\section{Introduction}

%
%
Many real-world imperfections arise when experimentally performing a
quantum information processing task.  These may arise either in the
creation or measurement of a quantum state, or in the manipulation of
the state via some quantum process. It is important to quantitatively
measure and characterize these imperfections in a way that is
theoretically meaningful and experimentally practical.

%
%
How can this be done?  Quantum states can be completely determined
using quantum state tomography~\cite{Jones91a,Leonhardt96a} and
compared using a variety of well-known measures~\cite{mikenike}.
Quantum processes can be measured using an analogous procedure called
quantum process tomography~\cite{Chuang97a,Poyatos97a,mikenike}.
%
%
However, the problem of developing quantitative measures to compare
real and idealized quantum processes has not been comprehensively
addressed.

Ideally there would be a single good measure, a ``gold
standard''~\cite{footnote1}, enabling
sensible comparison of different experimental implementations of
quantum information processing, and agreed upon by experimentalists
and theorists alike.  We will refer to candidates for such a gold
standard as ``distance measures'' for quantum processes, or as ``error
measures'', when we want to stress the comparison of real and
idealized processes.

%
%
Such an error measure would be extremely useful both when comparing
experiments with the theoretical ideal, and in comparing different
experiments that attempt to perform the same task.  Existing
experiments in quantum information processing have typically been
assessed on a rather \emph{ad hoc} basis.  For example, some
implementations of quantum logic gates have relied on demonstrating
that those gates act in the correct way on computational basis states
(i.e., verifying the truth table of the gate), and a few
superposition states.  Such demonstrations are important, but it is
clear that a figure of merit that is standardized, theoretically well
motivated and experimentally practical would be a considerable step
forward.  Parenthetically, we note that such a measure would also be
of great use in concretely connecting real experiments to results such
as the fault-tolerance threshold for quantum computation~\cite{footnote2}.

%
%
The purpose of this paper is to comprehensively address the problem of
developing such error measures.  There is a sizeable previous
literature on this subject, but we believe that there has been a
consistent gap between work motivated primarily by theoretical
considerations, and work constrained by experimental realities.  Our
paper aims to address both theoretical and experimental desiderata.

%
%
The key to our work is to introduce a list of six simple,
physically motivated criteria that should be satisfied by any good
measure of distance between quantum processes.  These criteria enable
us to eliminate many approaches to the definition of an error measure
that \emph{a priori} appear highly plausible.

%
%
The criteria are as follows.  Suppose $\Delta$ is a candidate measure
of the distance between two quantum processes.  Such processes are
described by maps between input and output quantum states, e.g.,
$\rho_{\text{out}}=\mathcal{E}(\rho_{\text{in}})$, where the map
${\cal E}$ is known as a \emph{quantum operation}~\cite{footnote3,mikenike}.  Physically,
$\Delta({\cal E},{\cal F})$ may be thought of in two ways: as a
measure of error in quantum information processing when one wants to
do the ideal process ${\cal F}$ but does ${\cal E}$ instead; or of
distinguishability between the two processes ${\cal E}$ and ${\cal
  F}$.  We believe that any such measure must satisfy the following
six properties, motivated by both physical and mathematical concerns.
  
(1) \emph{Metric:} $\Delta$ should be a metric.
  This requires three properties: (i) $\Delta({\cal E},{\cal F}) \ge
  0$ with $\Delta({\cal E},{\cal F}) =0$ if and only if
  $\mathcal{E}=\mathcal{F}$; (ii) Symmetry:
  $\Delta(\mathcal{E},\mathcal{F})=\Delta(\mathcal{F},\mathcal{E})$;
  and (iii) the triangle inequality $\Delta(\mathcal{E},\mathcal{G})\le
  \Delta(\mathcal{E},\mathcal{F})+\Delta(\mathcal{F},\mathcal{G})$.
  
(2) \emph{Easy to calculate:} it should be
  possible to evaluate $\Delta$ in a direct manner.
  
(3) \emph{Easy to measure:} there should be a
  clear and achievable experimental procedure for determining the
  value of $\Delta$.
  
(4) \emph{Physical Interpretation:} $\Delta$
  should have a well-motivated physical interpretation.
 
(5) \emph{Stability}~\cite{Aharonov98a}:
  $\Delta(\mathcal{I}\otimes\mathcal{E},
  \mathcal{I}\otimes\mathcal{F}) = \Delta(\mathcal{E},\mathcal{F})$,
  where ${\cal I}$ represents the identity operation on an additional
  quantum system.  Physically, this means that unrelated
  ancillary quantum systems do not affect the value of $\Delta$.
  
(6) \emph{Chaining:}
  $\Delta(\mathcal{E}_2\circ\mathcal{E}_1,
  \mathcal{F}_2\circ\mathcal{F}_1)\le
  \Delta(\mathcal{E}_1,\mathcal{F}_1)+\Delta(\mathcal{E}_2,\mathcal{F}_2)$.
  Thus, for a process composed of many smaller steps, the total
  error will be less than the sum of the errors in the individual steps.

%
%
The chaining and stability criteria are key properties for estimating
the error in a complex quantum information processing task.  Because
quantum information processing tasks are typically broken down into a
sequence of simpler component operations, a conservative bound on the
total error can be found by simply analyzing the individual
components.  This is critical for applications such as quantum
computation, where full process tomography on an $n$-qubit computation
requires exponentially many measurements, and is thus infeasible.
Chaining and stability enable one to instead benchmark the constituent
processes involved in the computation, which can then be used to infer
that the entire computation is robust.

%
%
Many other properties follow from these six criteria.  For example,
from the metric and chaining criteria we see that
$\Delta(\mathcal{R}\circ \mathcal{E},\mathcal{R}\circ \mathcal{F}) \le
\Delta(\mathcal{E},\mathcal{F})$, where ${\cal R}$ is any quantum
operation.  This corresponds to the requirement that post-processing
by $\mathcal{R}$ cannot \emph{increase} the distinguishability of two
processes $\mathcal{E}$ and $\mathcal{F}$.  Another elementary
consequence of the metric and chaining criteria is \emph{unitary
  invariance}, i.e., $\Delta({\cal U} \circ {\cal E} \circ {\cal
  V},{\cal U} \circ {\cal F} \circ {\cal V}) = \Delta({\cal E},{\cal
  F})$, where ${\cal U}$ and ${\cal V}$ are unitary operations.

%
%
For both theoreticians and experimentalists, there are strong
motivations to find a gold standard satisfying these criteria---the
need for a physically sensible way of evaluating the performance of a
quantum process, and the need to compare the success of a theoretical
model to the operation of a real, experimental system. For the
experimentalist, however, there is also another important
consideration. That is the need for \emph{diagnostic measures} which
can be used to build insight into the source of imperfections in
experimental implementations.  Diagnostic measures may not necessarily
be good candidates for our sought-after gold standard --- they may
fail to satisfy one or more of our criteria --- but they still may be
extremely useful in the experimental context.  Thus, some of the
measures we discard as unsuitable for use as a gold standard may still
be useful as diagnostic measures.  Furthermore, it is not difficult to
construct other examples of useful diagnostic measures, different to
any considered in this paper.  The detailed investigation of such
diagnostic measures is, however, beyond the scope of the present
paper.

%
%
\textit{Prior work:} The principal contribution of our paper is to
comprehensively evaluate many plausible error measures for quantum
information processing, within the broad framework of the criteria we
have identified.  So far as we are aware, none of the prior work has
surveyed and compared error measures against such a broad array of
theoretical and experimental concerns.

Error measures for quantum teleportation have received particular
attention in the prior literature, perhaps spurred by controversy over
which experiments should be regarded as definitively demonstrating the
teleportation effect~\cite{Bennett93a}.  Examples of this line of
development
include~\cite{Caves04a,Braunstein01a,Rudolph01a,Grosshans01a,Ralph99a,Schack99a},
and references therein.  With the exception of Ref.~\cite{Schack99a} this
work differs from ours in that it is focused primarily on the problem
of teleportation.  Reference\cite{Schack99a} has a more general focus, but is
not primarily concerned with the development of error measures, but
rather with the question of when quantum information processing can be
modeled classically.

More mathematical investigations of error measures have also been
mounted, especially in the context of quantum communication and
fault-tolerant quantum computation.  Examples of this work
include~\cite{Terhal04a,02nielsen249,Bowdrey02a,01raginsky11,Childs01a,Childs00a,Aharonov98a,Bernstein97a,Schumacher96a,Schumacher95a},
and references therein.  This work (often embedded in some larger
investigation) typically focuses on one or a few measures of specific
interest for the problem at hand.  These papers thus differ from our
work in that they don't attempt a comprehensive survey of possible
error measures against some set of abstract criteria; nor, typically,
do they address experimental criteria such as ease of measurement.
Nonetheless, while this prior work is different in character from
ours, it has greatly informed our point of view, and we will have
occasion to cite it on specific points throughout this paper.  Of
particular relevance is Ref.~\cite{Aharonov98a}, which introduced one of
the key measures we use, the stabilized process distance, or
\emph{S}~distance (referred to as the diamond norm in Ref.~\cite{Aharonov98a}),
and emphasized some of the important properties satisfied by that
measure.

%
%
\textit{Structure of the paper:} Secs.~\ref{sec:quantum-processes}
and~\ref{sec:distance-states} summarize background material on quantum
operations and distance measures for quantum states.

Section~\ref{sec:survey} is the core of the paper, comprehensively
surveying possible approaches to the definition of error measures.
Our strategy is to cast a wide net, considering many different
possible approaches to the definition of a distance measure, and then
to use our list of criteria to eliminate as many approaches as
possible.  This means a certain amount of tedium as we propose and
then reject certain \emph{a priori} plausible candidate error
measures.  The benefit of going through this process of elimination is
considerable, however.  First, it gives us confidence that the few
measures we identify as particularly promising should be preferred
over all other measures.
Indeed, we quickly eliminate all but four of
the measures we define as follows: the \emph{Jamiolkowski process fidelity}
(\emph{J}~fidelity), the \emph{Jamiolkowski process distance} (\emph{J}~distance),
the \emph{stabilized process fidelity} (\emph{S}~fidelity), and the
\emph{stabilized process distance} (\emph{S}~distance).  Second, in several
instances we show that error measures proposed previously in the
literature (in one case, by one of the authors of this paper) should
be rejected as inadequate.

Section~\ref{sec:application} applies the four promising measures
identified in Sec.~\ref{sec:survey} to the concrete problem of quantum
computation, showing that each measure has a useful operational
interpretation in terms of the success or failure of a quantum
computation.  

%
%
Section~\ref{sec:conclusion} concludes the paper with a summary of our
results, and the identification of the \emph{S}~distance and the \emph{S}~fidelity
as the two measures whose properties make them the most attractive
candidates for use as a gold standard in quantum information
processing.  We do not make a final recommendation as to which of
these two measures should be used, since they have extremely similar
strengths and weaknesses.  However, we do discuss and make definite
recommendations regarding the reporting of quantum information
processing experiments.  Furthermore, we sketch future research
directions which may ameliorate some of the weaknesses of one or both
measures, and which may therefore make it possible to definitively
choose a single measure as a gold standard.

\section{Describing quantum processes}
\label{sec:quantum-processes}

%
%
Quantum operations describe the most general physical processes that
may occur in a quantum system~\cite{mikenike}, including unitary
evolution, measurement, noise, and decoherence.  Any quantum operation
may be given the \emph{operator-sum representation} relating input
$\rho_{\text{in}}$ and output $\rho_{\text{out}}$ states,
\begin{equation}\label{eq:op-sum} 
\rho_{\text{out}} = \mathcal{E}(\rho_{\text{in}}) = \sum_j 
E_j \rho_{\text{in}} E_j^\dagger,
\end{equation} 
where the operators $E_j$ are known as \emph{operation elements}, and
obey the condition that
$\sum_j E_j^\dagger E_j\le I$~\cite{footnote4}.  Note that the operation elements $\{
E_j\}$ completely describe the effect of the process.  We will mostly
be concerned with the case of trace-preserving operations, for which
$\sum_j E_j^\dagger E_j = I$.  Physically, this corresponds to the
requirement that ${\cal E}$ represents a physical process without
post-selection~\cite{footnote5}.  Many of our results
extend easily to the case of non trace-preserving operations, but to
ease the exposition we assume processes are trace-preserving unless
otherwise noted.

%
%
The operator-sum representation has the drawback that it is not
unique, in the sense that there is a freedom in the choice of
operation elements~\cite{mikenike}. This is inconvenient if we are
trying to compare two processes. To alleviate this, let us fix a basis
$\{A_j\}$ for the space of operators, choosing for convenience a
basis orthonormal under the Hilbert-Schmidt inner product,
i.e., $\mbox{tr}(A_j^\dagger A_k) = \delta_{jk}$~\cite{footnote6}.
We can use this basis to expand the operation elements, $E_j = \sum_m
a_{jm} A_m$, and rewrite Eq.~(\ref{eq:op-sum}):
\begin{equation}
    \label{eq:op-in-basis}
    \mathcal{E}(\rho) = \sum_{mn} (\chi_\mathcal{E})_{mn} A_m \rho
    A_n^\dagger
\end{equation}
where $(\chi_\mathcal{E})_{mn}\equiv \sum_j a_{jm}a_{jn}^*$ are the
elements of the \emph{process matrix}, $\chi_{\cal E}$.
Equation~(\ref{eq:op-in-basis}) tells us that the process matrix completely
describes the action of the quantum process.  The big advantage of the
process matrix representation is that, unlike the operator-sum
representation, once the basis $\{ A_j\}$ is chosen the process matrix
can be shown to be unique to the process~\cite{footnote7};
i.e., it depends only on ${\cal E}$, not on the particular choice of
operation elements $\{ E_j\}$.  We will not give an explicit proof of
this fact here, but note that this result follows easily from the
discussion below.

%
%
The process matrix gives a convenient way of representing the
operation ${\cal E}$.  A closely-related but more abstract
representation is provided by the \emph{Jamiolkowski
  isomorphism}~\cite{72jamiolkowski275}, which relates a quantum
operation ${\cal E}$ to a quantum state, $\rho_{\cal E}$:
\begin{eqnarray}
\label{eq:isomorphism}
  \rho_{\cal E} 
\equiv [\mathcal{I} \otimes \mathcal{E}](\ketbra{\Phi}{\Phi}),
\end{eqnarray}
where $\ket{\Phi}=\sum_j\ket{j}\ket{j}/\sqrt{d}$ is a
maximally entangled state of the ($d$-dimensional) system with another
copy of itself, and $\{\ket{j}\}$ is some orthonormal basis set.  The
map ${\cal E} \rightarrow \rho_{\cal E}$ is invertible,
that is, knowledge of $\rho_{\cal E}$
is equivalent to knowledge of ${\cal E}$~\cite{footnote8}.  This isomorphism thus
allows us to treat quantum operations using the same tools as are
ordinarily used to treat quantum states.  For later use we note the
useful property $\rho_{\mathcal{E}\otimes\mathcal{F}} =
\rho_{\mathcal{E}}\otimes\rho_{\mathcal{F}}$.

%
%
The state $\rho_{\cal E}$ and the process matrix $\chi_{\cal E}$ are
closely related.  A direct calculation shows that if one chooses the
operator basis sets $\{ A_j\} =\{ |m\rangle \langle n|\}$, then
$\chi_{\cal E} = d \rho_{\cal E}$, as matrices.  Thus we shall refer
to both $\chi_{\cal E}$ and $\rho_{\cal E}$ as the process matrix, and
treat them interchangeably.  This is very convenient, as $\rho_{\cal
  E}$ is easy to work with mathematically, using the expression
Eq.~(\ref{eq:isomorphism}), while the elements of $\chi_{\cal E}$ have
an obvious physical significance, expressed by
Eq.~(\ref{eq:op-in-basis}).

%
%
We conclude this section with a comment on our notational conventions.
We often use notation like $\psi$ to denote either a pure state
$|\psi\rangle$ or the corresponding density matrix $|\psi\rangle
\langle \psi|$, with the meaning to be determined from context.  Thus,
for example, we may write $\psi = \alpha |0\rangle +\beta|1\rangle$ to
indicate a pure state of a single qubit, while also writing ${\cal
  E}(\psi)$ to indicate a quantum
operation ${\cal E}$ acting on the density matrix corresponding to  
that pure state.

\section{Distance measures for quantum states} 
\label{sec:distance-states}

%
%
A natural starting place for an attempt to define a measure of
distance for quantum processes is measures of distance for quantum
states.  The quantum information science community has identified the
\emph{trace distance} and the \emph{fidelity} as particularly
important approaches to the definition of a distance measure for
states~\cite{footnote9}, and these two
measures will serve as the basis for our later definitions of distance
measures for quantum operations.  In keeping with the aims of the
paper, we don't make a choice between the trace distance and the
fidelity at the outset. Instead, our preference is to develop distance
measures for quantum operations based on \emph{both} the trace
distance and the fidelity, and then assess them using the criteria
discussed in the introduction.  We now briefly review the basic
properties of the trace distance and the fidelity.

%
%
\textit{The trace distance:} The \emph{trace distance} between density
matrices $\rho$ and $\sigma$ is defined by $D(\rho,\sigma) \equiv
\frac 12 \mbox{tr}|\rho-\sigma|$, where $|X| \equiv \sqrt{X^\dagger
  X}$.  From this definition it follows that the trace distance is a
genuine metric on quantum states, with $0\le D\le1$.  The trace
distance also has many other attractive properties that make it a
particularly good measure of distance between quantum states.  We now
briefly describe three of these.

%
%
First, the trace distance has a compelling physical interpretation as
a measure of state distinguishability.  Suppose Alice prepares a
quantum system in the state $\rho$ with probability $\frac 12$, and in
the state $\sigma$ with probability $\frac 12$.  She gives the system
to Bob, who performs a POVM measurement~\cite{mikenike} to distinguish
the two states.  It can be shown that Bob's probability of correctly
identifying which state Alice prepared is $1/2 + D(\rho,\sigma)/2$.
That is, $D(\rho,\sigma)$ can be interpreted, up to the factor 1/2, as
the optimal \emph{bias} in favour of Bob correctly determining which
of the two states was prepared.  This physical interpretation follows
from the identity
$D(\rho,\sigma) = \max_{E \leq I} \mbox{tr}(E(\rho-\sigma))$~\cite{footnote10}, where
the maximum is over all positive operators $E$ satisfying $E \leq I$.

%
%
Second, the trace distance possesses the \emph{contractivity}
property~\cite{Ruskai94a}, that is, $D({\cal E}(\rho),{\cal
  E}(\sigma)) \leq D(\rho,\sigma)$ whenever ${\cal E}$ is a
trace-preserving quantum operation.  This statement expresses the
physical fact that a quantum process acting on two quantum states
cannot increase their distinguishability.  Contractivity follows from
the physical interpretation of $D(\rho,\sigma)$ described above.

%
%
Third, the trace distance is \emph{doubly convex}, i.e., if $p_j$ are
probabilities then $D(\sum_j p_j \rho_j, \sum_j p_j \sigma_j) \leq
\sum_j p_j D(\rho_j,\sigma_j)$.  This inequality can be physically
interpreted as the statement that the distinguishability between the
states $\sum_j p_j \rho_j$ and $\sum_j p_j \sigma_j$, where $j$ is not
known, can never be greater than the average distinguishability when
$j$ is known, but has been chosen at random according to the
distribution $p_j$.

%
%
\textit{Fidelity:} The \emph{fidelity} between density matrices $\rho$
and $\sigma$ is defined by
\begin{eqnarray} \label{eq:fidelity}
F(\rho,\sigma) \equiv \mbox{tr}\left(
  \sqrt{\sqrt{\rho} \sigma \sqrt{\rho}}\right)^2. 
\end{eqnarray}
When $\rho = \psi$ is a pure state, this reduces
to $F(\psi,\sigma) = \langle \psi|\sigma|\psi\rangle$, the overlap
between $\psi$ and $\sigma$.

%
%
The fidelity also has many attractive properties.  It can be shown
that $0 \leq F(\rho,\sigma) \leq 1$, with equality in the second
inequality if and only if $\rho = \sigma$.  The fidelity is thus not a
metric as such, but serves rather as a generalized measure of the
overlap between two quantum states.  The fidelity is also symmetric in
its inputs, $F(\rho,\sigma) = F(\sigma,\rho)$, a fact that is not
obvious from the definition we have given, but which follows from
other equivalent definitions.

%
%
There is an ambiguity in the literature in the definition of fidelity
that is worth commenting on here.  Both the quantity defined above and
its square root have been referred to as the fidelity, and
both have many appealing properties~\cite{footnote11}.

%
%
Nevertheless, we strongly advocate using the definition of
Eq.~(\ref{eq:fidelity}), despite the other definition being used in
references such as~\cite{mikenike}.  As we will see in
Sec.~\ref{sec:application}, adopting the definition of
Eq.~(\ref{eq:fidelity}) gives rise to a measure of distance between
quantum processes with a physically compelling interpretation in terms
of the \emph{probability of success} of a quantum computation.
Adopting the other definition of fidelity would make about as much
sense as reporting the square root of the probability that the quantum
computation succeeded.

%
%
Although not a metric, the fidelity can easily be turned into a
metric.  Two common ways of doing this are the \emph{Bures metric},
defined by $B(\rho,\sigma) \equiv \sqrt{2- 2\sqrt{F(\rho,\sigma)}}$,
and the \emph{angle}, defined by $A(\rho,\sigma) \equiv \arccos
\sqrt{F(\rho,\sigma)}$.  The origin of these metrics can be seen
intuitively by considering the case when $\rho$ and $\sigma$ are both
pure states.  The Bures metric is just the Euclidean distance between
the two pure states, with respect to the usual norm on state
space~\cite{footnote12}, while the angle is, as the name
suggests, just the angle between the two states, with respect to the
usual inner product on state space.

%
%
In addition to the angle and the Bures metric we will find it
convenient to introduce a third metric based on the fidelity.  This
metric does not seem to have been previously recognized in the
literature, but arises naturally later in this paper in the context of
quantum computation. It is defined by $C(\rho,\sigma) \equiv
\sqrt{1-F(\rho,\sigma)}$.  The only difficult step in proving this is
a metric is the proof of the triangle inequality~\cite{footnote13}.

%
%
In later sections our discussion will sometimes focus on the fidelity,
and sometimes on metrics derived from the fidelity.  We will say that
a metric $\Delta^F(\rho,\sigma)$ on state space is a
\emph{fidelity-based} metric if it is a monotonically decreasing
function of the fidelity $F(\rho,\sigma)$.  Obviously the angle, the
Bures metric and $C(\cdot,\cdot)$ are all fidelity-based metrics.  It
is often the case that the specific details of the metric used are not
important, and whenever possible we state results using the fidelity
as a single unifying concept.  However, sometimes it will prove
advantageous to use the fidelity-based metrics directly.  In
particular, they have the advantage of satisfying the triangle
inequality, which turns out to be useful proving the chaining
criterion [property~(6)].

%
%
Like the trace distance, the fidelity and its derived metrics
have many other nice properties.  It can be shown~\cite{Barnum96a}
that $F({\cal E}(\rho),{\cal E}(\sigma)) \geq F(\rho,\sigma)$ for any
trace-preserving quantum operation ${\cal E}$.  We call this the
\emph{monotonicity} property of the fidelity.  It follows that any
fidelity-based metric satisfies a contractivity property analogous to
that satisfied by the trace distance.

%
%
The fidelity also satisfies a property analogous to the double
convexity of the trace distance.  Precisely, the square root of the
fidelity is \emph{doubly concave}, that is, $F(\sum_j p_j
\rho_j,\sum_j p_j \sigma_j)^{1/2} \geq \sum_j p_j
F(\rho_j,\sigma_j)^{1/2}$.  This double concavity can be used to prove
double convexity of certain fidelity-based metrics.  In particular,
supposing $\Delta^F$ is a fidelity-based metric which is convex in the
square root of the fidelity (the angle, the Bures metric and
$C(\cdot,\cdot)$ are all easily verified to have this property), then
it is easy to verify that $\Delta^F$ is doubly convex.

%
%
One drawback of the fidelity is that it is difficult to find a
compelling physical interpretation.  When $\rho$ and $\sigma$ are
mixed states, no completely satisfactory interpretation of the
fidelity is known (but c.f. Refs~\cite{Dodd02b,Fuchs96a}).  When $\rho =
\psi$ is a pure state, we have $F(\psi,\sigma) = \langle
\psi|\sigma|\psi\rangle$, the overlap between $\psi$ and $\sigma$.
Physically, we might imagine $\sigma$ is an attempt to prepare the
pure state $\psi$.  In this case the fidelity coincides with the
probability that a perfect measurement testing whether the state is
$\psi$ will succeed.  It is this property of the fidelity that is used
in Sec.~\ref{sec:application} to connect our fidelity-based error
measures for quantum processes to the probability of success of a
quantum computation.

\textit{General comments:} The fidelity is, at present, perhaps
somewhat more widely used in the quantum information science community
than is the trace distance.  However, we shall see below that the
trace distance and the fidelity have complementary advantages as a
basis for developing measures of distance for quantum operations, and
so it is useful to investigate both.  In any case, the two measures
are, as one might expect, quite closely related.  In particular, it is
possible to show that they are related by the
inequalities~\cite{Fuchs99a}:
\begin{eqnarray} \label{eq:relationship-of-measures}
  1-\sqrt{F(\rho,\sigma)} \leq D(\rho,\sigma) \leq \sqrt{1-F(\rho,\sigma)}.
\end{eqnarray}
It is not difficult to construct examples of saturation for both
inequalities.  Note that the second inequality is always saturated for
pure states, i.e., $D(\psi,\phi) = \sqrt{1-F(\psi,\phi)}$ for pure
states $\psi$ and $\phi$.

\section{Error measures for quantum processes}
\label{sec:survey}

Our goal in this paper is to recommend a single error measure enabling
researchers to compare the performance of quantum information
processing experiments against the theoretical ideal.  As the basis
for such a recommendation, in this section we comprehensively survey
possible definitions of such error measures, and do a preliminary
assessment of each measure against the criteria introduced earlier in
this paper.

%
%
We take three basic approaches to defining an error measure for
processes.  In Sec.~\ref{subsec:Jamiolkowski} we investigate
approaches based on the process matrix, $\rho_{\cal E}$.  In
Sec.~\ref{subsec:average-case} we investigate approaches based on
the \emph{average} behaviour of a process.  Finally, in
Sec.~\ref{subsec:worst-case} we investigate approaches based on the
\emph{worst-case} behaviour of a process.  In each case we investigate
measures based on both the trace distance and the fidelity.  We will
describe connections between the various measures, and identify four
measures of particular merit.  The properties of these four measures
will be discussed in more detail in the next section.

%
%
\textit{Nomenclature:} In the following treatment we shall use the
unadorned symbol $\Delta$ to mean a metric between states. Our
approach is to use state-based metrics to form metrics between
processes, and these will also be represented by $\Delta$ but with a
subscript denoting the method used, e.g. $\Delta_{\rm ave}$ is a
process metric based on the average over input states. Where we need
to specialize to a specific state-metric we will use a superscript
with the symbol representing that metric ($A$, $B$, $C$, and $D$ from
section~\ref{sec:distance-states}), or use that symbol directly with a
subscript for the method, e.g.  $\Delta_{\mathrm{ave}}^{D}\equiv
D_{\mathrm{ave}}$ is the \emph{process} metric based on the average
trace distance. The chief departure from these conventions will be due
to the fidelity, which is not a metric. We will use the notation
$\Delta^F$ to mean any \emph{metric} derived from the fidelity (e.g.
$A$, $B$, and $C$) and the symbol $F$ with a subscript to mean a
process measure based on fidelity, for example $F_\mathrm{ave}$ is the
average fidelity.

\subsection{Error measures based on the process matrix}
\label{subsec:Jamiolkowski}

Suppose $\Delta(\rho,\sigma)$ is any metric on the space of quantum
states.  A natural approach to defining a measure $\Delta_{\rm pro}$
of the distance between two quantum processes is
\begin{equation}
\Delta_{\rm pro}({\cal E},{\cal F}) \equiv \Delta(\rho_{\cal E},\rho_{\cal F}). 
\label{eqn:Delta_pro}
\end{equation}
%
%
Defining $\Delta_{\rm pro}$ in this way automatically gives
$\Delta_{\rm pro}$ the metric property. Provided $\Delta(\cdot,\cdot)$
is easy to calculate, $\Delta_{\rm pro}$ is also easy to calculate.
Furthermore, since ${\cal E}$ can be experimentally determined using
quantum process tomography, it follows that $\Delta_{\rm pro}$ can be
experimentally measured, at least in principle.

%
%
What about the other properties?  The properties of stability and
chaining can be obtained by making some natural extra assumptions
about the state metric $\Delta$, which we now describe.  Suppose first
that the metric $\Delta$ is \emph{stable} in the sense that
$\Delta(\rho \otimes \tau, \sigma \otimes \tau) =
\Delta(\rho,\sigma)$.  This is easily seen to be the case for the
trace distance and for any fidelity-based metric, for example.  The
stability property for $\Delta_{\rm pro}$ follows immediately: \\
$\Delta_{\rm pro}(I\otimes\mathcal{E},I\otimes\mathcal{F})=
\Delta(\rho_I\otimes \rho_\mathcal{E},\rho_I \otimes
\rho_\mathcal{F})= \Delta(\rho_{\cal E}, \rho_{\cal F}) = \Delta_{\rm
  pro}(\mathcal{E},\mathcal{F})$.

%
%
The chaining property can be proved, with some caveats to be described
below, by assuming that $\Delta(\cdot,\cdot)$ is \emph{contractive}, i.e.,
$\Delta({\cal E}(\rho),{\cal E}(\sigma)) \leq \Delta(\rho,\sigma)$, for
trace-preserving operations ${\cal E}$.  We have already seen that this
is a natural physical assumption satisfied by the trace distance and
any fidelity-based metric.

%
%
Suppose then that $\Delta$ is contractive with respect to
trace-preserving operations.  We claim that $\Delta_{\rm pro}$
satisfies the chaining property, \\
$\Delta_{\rm pro}({\cal E}_2 \circ
{\cal E}_1,{\cal F}_2 \circ {\cal F}_1) \leq \Delta_{\rm pro}({\cal
  E}_2,{\cal F}_2)+\Delta_{\rm pro}({\cal E}_1,{\cal F}_1)$, \\
  provided
${\cal F}_1$ is \emph{doubly stochastic}, i.e., ${\cal F}_1$ is
trace-preserving and satisfies ${\cal F}_1(I) = I$; this assumption is
used at a certain point in our proof of chaining.  This may seem like
a significant assumption, since physical processes such as relaxation
to a finite temperature are not doubly stochastic.  However, in
quantum information science we are typically interested in the case
when ${\cal F}_1$ and ${\cal F}_2$ are ideal unitary processes, and we
are using $\Delta_{\rm pro}$ to compare the composition of these two
ideal processes to the experimentally realized process ${\cal E}_2
\circ {\cal E}_1$.  Since unitary processes are automatically doubly
stochastic, it follows that chaining holds in this case, which is the
case of usual interest.

%
%
The proof of chaining begins by applying the triangle inequality to obtain
\begin{eqnarray}
\Delta_{\rm pro}({\cal E}_2 \circ {\cal E}_1,{\cal F}_2 \circ {\cal F}_1) 
& = &
\Delta(\rho_{{\cal E}_2 \circ {\cal E}_1},\rho_{{\cal F}_2 \circ {\cal
    F}_1}) \\
& \leq & \Delta(\rho_{{\cal E}_2 \circ {\cal E}_1},\rho_{{\cal E}_2 
    \circ {\cal F}_1})  \nonumber \\
& & + \Delta(\rho_{{\cal E}_2 \circ {\cal F}_1},
  \rho_{{\cal F}_2 \circ {\cal F}_1}). \label{eq:chaining-intermediate}
\end{eqnarray}
Then note the easily-verified identity
$\rho_{\mathcal{E}\circ\mathcal{F}}= (\mathcal{F}^T \otimes
\mathcal{E})(\Phi)$, where $\Phi$ is the maximally entangled state
defined earlier, we define $\mathcal{F}^T(\rho)\equiv \sum_j F_j^T\rho
F_j^{*}$, and $F_j$ are the operation elements for ${\cal F}$ [c.f.
Eq.~(\ref{eq:op-sum})].  Applying this identity to both density
matrices in the second term on the right-hand side of
Eq.~(\ref{eq:chaining-intermediate}) gives
\begin{eqnarray}
& & \Delta_{\rm pro}({\cal E}_2 \circ {\cal E}_1,{\cal F}_2 \circ {\cal F}_1) 
\nonumber \\
& \leq & \Delta(\rho_{{\cal E}_2 \circ {\cal E}_1},\rho_{{\cal E}_2 
    \circ {\cal F}_1})  \nonumber \\
& & + \Delta(({\cal F}_1^T \otimes {\cal E}_2)(\Phi),
  ({\cal F}_1^T \otimes {\cal F}_2)(\Phi)). \label{eq:chaining-inter-2}
\end{eqnarray}
The double stochasticity of ${\cal F}_1$ implies that ${\cal F}_1^T$
is a trace-preserving quantum operation.  We can therefore apply
contractivity to both the first and the second terms on the right-hand
side of Eq.~(\ref{eq:chaining-inter-2}), giving the desired result.

%
%
Only one property of $\Delta_{\rm pro}$ remains in question, and that is whether
or not it has a good physical interpretation.  We will see in
Sec.~\ref{sec:application} that $D_{\rm pro}$ and $F_{\rm pro}$ can
both be related in a natural way to the average probability with which
a quantum computation fails or succeeds, providing a good physical
interpretation for these quantities.

%
%
Although $\Delta_{\rm pro}$ may be calculated easily in principle for
both the trace distance and fidelity-based approaches, the
fidelity-based measures have some substantial advantages.  The reason
is that, so far as we are aware, experimentally determining $D_{\rm
  pro}$ requires doing full process tomography, which for a
$d$-dimensional quantum system requires the estimation of $d^4-d^2$
observable averages.  By contrast, when $U$ is a unitary operation it
turns out that the fidelity $F_{\rm pro}({\cal E},U)$ (and related
error measures) can be determined based upon the estimation of at most 
$2d^2$ observable averages, and in particular, $d^2$ observable averages for qubits.  
This makes $F_{\rm pro}({\cal E},U)$ and
related error measures substantially easier to determine
experimentally than $D_{\rm pro}$.  The key to proving this is the
observation~\cite{footnote14}
\begin{eqnarray} \label{eq:process-est}
  F_{\rm pro}({\cal E},U) = 
  \frac{1}{d^3} \sum_j \mbox{tr}(U U_j^\dagger U^\dagger
  {\cal E}(U_j)),
\end{eqnarray}
where the $\{ U_j \}$ are a basis of unitary operators orthogonal
under the Hilbert-Schmidt inner product, satisfying
$\mbox{tr}(U_j^\dagger U_k) = d \delta_{jk}$.  Up to scaling we saw an
example of such a set in Sec.~\ref{sec:quantum-processes}, the
$n$-qubit tensor products formed from the Pauli matrices and the
identity matrix.  Equation~(\ref{eq:process-est}) does not provide a direct
way of estimating $F_{\rm pro}$.  But suppose we expand the $U_j$ in
terms of a set of \emph{input states}, $\rho_k$: $U_j = \sum_k a_{jk}
\rho_k$.  These input states must span the entire operator space, and
thus there must be $d^2$ of them; we will see an explicit example below
for two qubits.  We also expand $U U_j U^\dagger$ in terms of a set of
\emph{observables}, $\sigma_l$: $U U_j U^\dagger = \sum_l b_{jl}
\sigma_l$.  These observables must also span the entire operator
space.  Substitution into Eq.~(\ref{eq:process-est}) gives
\begin{eqnarray} \label{eq:fidelity-evaluate}
  F_{\rm pro}({\cal E},U) = \frac{1}{d^3} \sum_{kl} M_{kl} 
  \mbox{tr}(\sigma_l{\cal E}(\rho_k)),
\end{eqnarray}
where $M_{kl} \equiv \sum_j b_{jl} a_{jk}$.  This equation gives a
method to evaluate $F_{\rm pro}$: choose a spanning set of $d^2$ input
states $\rho_k$ which can be prepared experimentally, and a set of
observables $\sigma_l$ whose averages we can reliably measure;
determine the matrix $M = (M_{kl})$, whose elements depend only on
known quantities ($\rho_k, \sigma_l$, and the idealized operation
$U$), not on the unknown ${\cal E}$.  The non-zero matrix elements in
$M$ will determine which observable averages need to be estimated for
calculating $F_{\rm pro}({\cal E},U)$.  In general, $d^4$ observable
averages will need to be estimated.  However, suppose we choose some
fixed set of $\rho_k$, and then define  $\sigma_l
\equiv \sum_k a_{kl} U U_k U^\dagger$~\cite{footnote15}.  In this case it is easily
verified that Eq.~(\ref{eq:fidelity-evaluate}) simplifies to:
\begin{eqnarray}\label{eq:fidelity-min}
  F_{\rm pro}({\cal E},U) = \frac{1}{d^3} \sum_{k} 
  \mbox{tr}(\sigma_k{\cal E}(\rho_k)),
\end{eqnarray}
which only requires between $d^2$ and $2d^2$
measurements.  The drawback is that in
this method we are not free to choose the $\sigma_l$; they are
determined by $U$ and the $\rho_k$.

%
%
In practical situations, certain input states and measurements are
easier to use than others. We envisage an experimentalist choosing the
set of input states and measurements according to convenience and
using the prescription above to calculate which combinations are
necessary. This in general will be less than what is required to
perform full process tomography.  This direct method has the
additional advantage of making it easier to estimate the experimental
error in $F_{\rm pro}$.

%
%
For example, consider an $n$-qubit process, $U$. Suppose we select the
$U_j$ to range over the $n$-fold tensor products of Pauli matrices
(including the identity matrix).  Suppose furthermore that for each
qubit we select the input states from the set $\{I,I+X,I+Y,I+Z\}$
(where $X$, $Y$, $Z$ are the usual Pauli operators), so that we choose
$\rho_k$ from the set of all possible tensor products of the single
qubit input states.  Now, choosing $\sigma_l \equiv \sum_k a_{kl} U
U_k U^\dagger$, we see that the $a_{kl}$ will always be real, and
since the $U_k$ are Hermitian then the $\sigma_l$ are also Hermitian.
Thus Eq.~(\ref{eq:fidelity-min}) tells us that we need to estimate
only $d^2$ observable averages to evaluate $F_{\rm pro}$ for
\emph{any} $U$, much fewer than the $d^{4}-d^{2}$ observable averages
necessary to do full process tomography on $n$ qubits.

%
%
It is an interesting problem deserving further exploration to find the
minimum number of measurements required to estimate $F_{\rm pro}$ when
there are constraints on what input states and observables are
available.  For instance, it would be useful to know the optimal
number for the case where we are restricted to separable inputs and
product observables, i.e., inputs and observables that can be given
direct local implementations.

\subsection{Error measures based on the average case}
\label{subsec:average-case}

Another natural approach for defining error measures for quantum
operations is to compare output states and average over all input
state, where the output states can be compared using the distance
measures for states described in Section~\ref{sec:distance-states}.
We define
\begin{eqnarray} \label{eq:avg}
  \Delta_{\rm ave}({\cal E},{\cal F}) \equiv \int d\psi \, \Delta({\cal E}(\psi),
  {\cal F}(\psi)),
\end{eqnarray}
where the integral is over the uniform (Haar) measure on state space.

While this approach seems intuitively sensible, it turns out that the
resulting measures satisfy few of our criteria. The only two
properties these measures appear to satisfy in general, for an
arbitrary state metric $\Delta$, are the metric and chaining criteria,
both of which follow immediately from the metric property of $\Delta$.

The average-based metrics are less successful in meeting the other
criteria. Even when $\Delta$ is easy to calculate, it is not obvious
that the integral in Eq.~(\ref{eq:avg}) will have a simple form that
enables easy calculation of $\Delta_{\rm ave}$.  This, in turn, means
that $\Delta_{\rm ave}$ may not be so easy to determine
experimentally.  So far as we are aware, no simple expressions are
known for $\Delta_{\rm ave}$ for any of the metrics we have discussed.

It is not surprising that the physical interpretations of these
metrics rely heavily on the possible interpretations of the
corresponding state metrics as discussed in section
\ref{sec:distance-states}. The earlier discussion of the trace
distance, for example, follows on to give a meaning for $D_{\rm ave}$.
Suppose we are asked to distinguish between ${\cal E}(\psi)$ and
${\cal F}(\psi)$ for some $\psi$ which is known, but has been chosen
uniformly at random.  On average, the optimal probability of
successfully distinguishing the two processes will be $1/2+ D_{\rm
  ave}({\cal E},{\cal F})/2$.  Thus, $D_{\rm ave}({\cal E},{\cal F})$
may be interpreted as a measure of the average bias in favour of
correctly distinguishing which process was applied to a state $\psi$.
With regard to the fidelity-based metrics, however, there does not
appear to be any clear physical interpretation for $\Delta_{\rm ave}$
because of the lack of any clear meaning for the fidelity-based
metrics.

Finally, completing the checklist of criteria, our numerical analysis
shows that $\Delta_{\rm ave}$ is not stable for any of the four
candidate state metrics we've investigated.  Later in the paper we
describe in detail a method for ``stabilizing'' measures which are not
stable; we now briefly note the results that are obtained when this
procedure is applied in the present context.  The idea is to introduce
an ancillary system $A$, and consider the quantity $\Delta_{{\rm
    stab-ave}}({\cal E},{\cal F}) \equiv \lim \Delta_{\rm ave}({\cal
  I} \otimes {\cal E},{\cal I} \otimes {\cal F})$, where the limit is
that of large ancilla dimension.  Using the well-known result that a
randomly chosen chosen state of a composite system $AQ$ ($\dim A \gg
\dim Q$) has very close to maximal entanglement~\cite{Page93a,Sen96a},
it follows that $\Delta_{{\rm stab-ave}}({\cal E},{\cal F}) =
\Delta_{\rm pro}({\cal E},{\cal F})$, i.e., the stabilized average
distance reduces to the process distance considered earlier.

%
%
There is an alternative approach, available because the fidelity-based
metrics are nonlinear functions of the fidelity, which is to create a
measure based on the average fidelity:
\begin{eqnarray}
  F_{\rm ave}({\cal E},{\cal F}) \equiv \int d\psi \, 
  F({\cal E}(\psi),{\cal F}(\psi)).
  \label{eq:ave-fid}
\end{eqnarray}
When ${\cal F}$ is a unitary operation, $U$, the average fidelity has
a physical interpretation that is at least plausible, as the average
overlap between $U|\psi\rangle$ and ${\cal E}(\psi)$.  It was shown
in Ref.~\cite{Horodecki99c} (see also Ref.~\cite{02nielsen249}) that $F_{\rm
  ave}$ and $F_{\rm pro}$ are related by the equation
\begin{eqnarray} \label{eq:average-equation}
  F_{\rm ave}({\cal E},U) = \frac{F_{\rm pro}({\cal E},U)d +1}{d+1}, 
\end{eqnarray}
where $d$ is the dimension of the quantum system, and we are
restricting ourselves to the case where $U$ is a unitary operation.
This relationship makes $F_{\rm ave}({\cal E},U)$ easy to
calculate~\cite{Bowdrey02a,02nielsen249} and also easy to measure
experimentally, using the techniques described in the previous
subsection for $F_{\rm pro}({\cal E},U)$.

%
%
Although $F_{\rm ave}$ has several advantages (ease of calculation,
ease of measurement, and a physical interpretation), the outlook for
the other criteria is not so good.  Not only is $F_{\rm ave}$ not a
metric, it is not stable either, a fact that follows from
Eq.~(\ref{eq:average-equation}) and the knowledge that $F_{\rm pro}$
is stable. The same argument shows that measures analogous to $A$,
$B$, and $C$ based on $F_{\rm ave}$ will also not be stable. We do not
know of any stable metrics that may be derived as a function of
$F_{\rm ave}$, and Eq.~(\ref{eq:average-equation}) renders any such
metrics equivalent in content to functions based on $F_\mathrm{pro}$
so the only reason to use them would be if they had better
characteristics.

To summarize the results of this section, they show that none of
the average-case error measures we have defined are particularly
attractive. However, these negative results are vital because these
approaches are all fairly natural solutions one might take to defining
a plausible error measure. It was therefore important to consider them
carefully before choosing to reject them.

\subsection{Error measures based on the worst case} 
\label{subsec:worst-case}

Our final approach to defining error measures is based on the worst
case distance between ${\cal E}(\psi)$ and ${\cal F}(\psi)$.  We
define
\begin{eqnarray}
  \Delta_{\max}({\cal E},{\cal F}) 
  \equiv \max_\psi \Delta({\cal E}(\psi),{\cal F}(\psi)),
\end{eqnarray}
where the maximum is over all possible pure state inputs, $\psi$, and
$\Delta$ is a metric on quantum states. 

When $\Delta =
\Delta^F$ is a fidelity-based metric, we see $\Delta_{\max}^{F}$ is a
function of the \emph{minimal fidelity}, defined by
\begin{eqnarray}
  F_{\min}({\cal E},{\cal F}) \equiv \min_\psi F({\cal E}(\psi),
  {\cal F}(\psi)).
\end{eqnarray}

%
%
In the definition of $\Delta_{\max}$, we maximize over all \emph{pure}
state inputs. Is this maximum the same if \emph{all} physical inputs,
including mixed states, are considered? In fact, it is fairly simple
to show that this is true, and therefore that it does not matter if we
optimize over pure or mixed states~\cite{footnote16}.
Suppose $\Delta$ is a \emph{doubly convex} metric, as are all the
metrics discussed in this paper (c.f.
Sec.~\ref{sec:distance-states}). If the maximum is achieved at some
mixed state, $\rho$, then we have $\Delta_{\max} = \Delta({\cal
  E}(\rho),{\cal F}(\rho))$. Expanding $\rho = \sum_j p_j \psi_j$ as a
mixture of pure states, and applying double convexity we see that the
maximum must also be attained at some pure state $\psi_j$. A similar
argument holds for $F_{\min}$, based on the double concavity of the
fidelity.

%
%
To assess the suitability of these measures, it is useful to first
note that $D_{\max}$ has already been shown in general not to be
stable~\cite{Aharonov98a}, and similar arguments can be made to extend
this to the fidelity-based measures. In Ref.~\cite{Aharonov98a}, Aharonov
\emph{et al.}  resolve this difficulty by constructing a variant of
$D_{\max}$ which is stable, but which otherwise has extremely similar
properties to $D_{\max}$. We now describe how this procedure can be
extended to define a stable version of $\Delta_{\max}$ for an
arbitrary state metric $\Delta$, and defer for the moment discussion
of the other criteria.

%
%
Suppose the original system $Q$ on which ${\cal E}$ and ${\cal F}$ act
has state space dimension $d$. It will be convenient to use subscripts
to indicate the system on which operations act (e.g. ${\cal E} = {\cal
  E}_Q, {\cal F} = {\cal F}_Q$). We introduce a fictitious
$d$-dimensional ancillary system $A$, acted on by the identity
operation ${\cal I}_A$, and define the stabilized
quantity~\cite{footnote17}
\begin{equation}
\Delta_{\rm stab}({\cal E}_Q,{\cal F}_Q) \equiv \Delta_{\max}({\cal
  I}_A \otimes {\cal E}_Q,{\cal I}_A \otimes {\cal F}_Q).
\end{equation}
The proof that $\Delta_{\rm stab}$ is stable under addition of systems
is simple and has been included in
Appendix~\ref{app:worst-case-stabilization}.  In the same way, we can
also define a stable form of the minimum fidelity,
$F_{\rm stab}({\cal E}_Q,{\cal F}_Q) \equiv F_{\min}({\cal I}_A\otimes {\cal E}_Q,{\cal
  I}_A \otimes {\cal F}_Q)$, with the proof of stability following
similar lines.  Note that the stabilized fidelity-based metrics
$\Delta^F_{\rm stab}$ are functions of $F_{\rm stab}$ in the obvious
way (e.g. we define as usual $A_{\rm stab}$, $B_{\rm stab}$ and
$C_{\rm stab}$).

%
%
Which of the other criteria for an error measure does $\Delta_{\rm
stab}$ satisfy? It is straightforward to show that $\Delta_{\rm stab}$
satisfies the metric and chaining criteria. Furthermore, the
stabilized trace-distance $D_{\rm stab}$ has an appealing physical
interpretation---it is the worst-case bias in the probability of being
able to distinguish $({\cal I} \otimes {\cal E})(\psi)$ from $({\cal
I} \otimes {\cal F})(\psi)$, where we allow an ancilla of arbitrary
size.  We defer discussion of the physical interpretation of the
fidelity-based measures until the next section, where we will see that
both they and $D_{\rm stab}$ can be given an elegant interpretation in
the context of quantum computation.

%
%
What of the remaining criteria, ease of calculation and ease of
measurement?  Unfortunately, no powerful general formulae for
calculating $\Delta_{\rm stab}$ are known.  Reference~\cite{Aharonov98a} gives a
general formula for the distance $D_{\rm stab}$ between two unitary
operations, but the more interesting case of the distance between an
idealized unitary operation and a noisy quantum process has not been
solved, even for single-qubit operations.

%
%
The good news is that $D_{\rm stab}$ and $F_{\rm stab}$ (and thus
$A_{\rm stab}, B_{\rm stab}$ and $C_{\rm stab}$) are easy to calculate
numerically, because they can all be reduced to \emph{convex
  optimization} problems \cite{convexopt}. For this special class of
problem, where the task is to minimize a convex function defined on a
convex set, extremely efficient numerical techniques are available.
Among many other nice properties, it is possible to show that a local
minimum of a convex optimization problem is always a global minimum,
and thus techniques such as gradient descent typically converge
extremely rapidly, with no danger of finding false minima. In
Appendix~\ref{app:proof-conv-opt}, we prove explicitly that finding
$F_{\rm stab}$ belongs to this class of problems, and the proof for
$D_{\rm stab}$ follows similar lines.

%
%
We have seen that numerical calculation of $D_{\rm stab}$ and $F_{\rm
  stab}$ can easily be carried out, and this enables a two-step
procedure for experimental measurement of either quantity---process
tomography, followed by a numerical optimization.  Of course, finding
general formulae along the lines of $F_{\rm pro}({\cal E},U)$ or
$D_{\rm pro}$ is still a highly desirable goal.  Aside from the
intrinsic benefit, finding general formulae would simplify the
experimental measurement and determination of error bars for $D_{\rm
  stab}$ and $F_{\rm stab}$, and perhaps obviate the need for a full
process tomography, as Eq.~(\ref{eq:process-est}) did for $F_{\rm
  pro}({\cal E},U)$.  

%
%
\section{Application to quantum computing} 
\label{sec:application}

Can we find a good physical interpretation for any of the error
measures that we've identified?  In this section we will focus on
interpretations that arise within the context of quantum computation
and we will find that of the error measures we have discussed, four
have particularly outstanding properties: $D_{\rm pro}$, $F_{\rm pro},
D_{\rm stab}$ and $F_{\rm stab}$.  (Note that in the case of the
fidelity, it will actually be more convenient to state our results in
terms of the equivalent measures $C_{\rm pro}$ and $C_{\rm stab}$.)

Assessed according to the criteria described in the introduction,
these four measures have already been found to be superior to all the
other measures we have studied.  The additional fact that each arises
naturally in the context of quantum computation strongly indicates
that these four measures are the most deserving of consideration as
measures of error in quantum information processing.  We will return
in the conclusion, Sec.~\ref{sec:conclusion}, to the question of which
of these four measures is the best possible measure of error.

%
%
There are a variety of different ways of describing quantum
computations, and it turns out that each of the four error measures
arises naturally in different contexts.  We will discuss separately
two broad divisions of quantum computation, \emph{function
  computation} and \emph{sampling computation} looking at both
worst-case and average-case performance for each division.

%
%
Most algorithms on classical computers are framed as function
computations.  We will see that our error measures can be given
particularly compelling interpretations relating to the probability of
error in a function computation.  However, in the context of
simulating quantum systems it is often more natural to consider
sampling computations, where the goal is to reproduce the statistics
obtained from a measurement of the system in some specified
configuration.  Again, we will see that our error measures can be
given good interpretations in this context, albeit somewhat more
complex interpretations than for function computation.

%
%
The reason for treating the two types of computation separately is at
least partially a practical one, since both types of computation arise
naturally in the context of quantum computation.  However, 
a more fundamental reason is that it does not appear to be
known how to reduce sampling computation to function computation.
Rather remarkably, even when there is an efficient way of
\emph{computing} a probability distribution, there does not
appear to be any general way to convert that into an efficient way of
\emph{sampling} from that distribution.

%
%
\subsection{Function computation}

In function computation, the goal of the quantum computation is to
\emph{compute a function}, $f$, exactly or with high probability of
success.  More precisely, the goal is to take as input an instance,
$x$, of the problem, and to produce a final state $\rho_x$ of the
computer that is either equal to $|f(x)\rangle$, or sufficiently close
that when a measurement in the computational basis is performed, the
outcome is $f(x)$ with high probability.  Grover's algorithm is
usually cast in this way, where we want to determine the identity of the
state marked by the oracle.

\textit{Function computation in the worst case:} Suppose we attempt to
perform a quantum computation represented by an ideal operation ${\cal
  F}$ that acts on an input $|x\rangle$, where $x$ represents the
instance of the problem to be solved, e.g., a number to be
factored~\cite{footnote18}.  This
process succeeds in computing $f(x)$ with an error probability of at
most $p_e^{\rm id}$, where `${\rm id}$' indicates that this is the
\emph{ideal} worst-case error probability.  Of course, in reality some
non-ideal operation ${\cal E}$ is performed.  A good measure of error
in the real computation is the \emph{actual} probability $p_e$ that
the measured output of the computation is not equal to $f(x)$. In
Appendix~\ref{app:function:worst}, we show that
\begin{eqnarray} \label{eq:fcsss1}
  p_e & \leq & p_e^{\rm id} + D_{\rm stab}({\cal E},{\cal F}) \\
\label{eq:fcsss2}
  p_e & \leq & \left[ \sqrt{p_e^{\rm id}} + C_{\rm stab}({\cal E},
        {\cal F}) \right]^2.
\end{eqnarray}
Which of these inequalities is better depends upon the exact
circumstances.  For example, when $p_e^{\rm id} = 0$, we see that
which inequality is better depends upon whether $D_{\rm stab}({\cal
  E},{\cal F})$ is larger or smaller than $C_{\rm stab}({\cal E},{\cal
  F})^2$.  With Eq.~(\ref{eq:relationship-of-measures}) in mind, it is
not difficult to convince oneself that either of these possibilities
may occur.

\textit{Function computation in the average case:} Once again our goal
is to compute a function $f(x)$ using an approximation ${\cal E}$ to
some ideal operation ${\cal F}$.  However, we now look at the
average-case error probability $\overline p_e$ that the measured
output of ${\cal E}(|x\rangle \langle x|)$ is not equal to $f(x)$,
where the average is taken with respect to a uniform distribution over
instances $x$.  Correspondingly, we introduce $\overline p_e^{\rm
  id}$, the average case error probability for the idealized operation
${\cal F}$.  We show that (App. \ref{app:function:average}):
\begin{eqnarray} 
  \label{eq:fcacss1}
  \overline p_e & \leq & \overline p_e^{\rm id} + D_{\rm pro}({\cal E},
  {\cal F}).
\end{eqnarray}
Unfortunately, we have been unable to develop a full natural analogue
of Eq.~(\ref{eq:fcsss2}) based on the fidelity. However, we have
proved a partial analogue for when the ideal computation succeeds with
probability one ($\overline p_e^{\rm id} = 0$). In this case:
\begin{eqnarray}   \label{eq:fcacss2}
  \overline p_e & \leq & C_{\rm pro}({\cal E},
        {\cal F})^2 = 1-F({\cal E},{\cal F}).
\end{eqnarray}
The proof uses very similar techniques to those used to establish
Eqs.~(\ref{eq:fcacss1}) and~(\ref{eq:fcsss2}), and is therefore
omitted.

%
%
\subsection{Sampling computation}

In sampling quantum computation, the goal is to \emph{sample} from
some ideal distribution $\{p_x(y)\} \equiv p_x$ on measurement
outcomes $y$, with $x$ representing input data for the problem.  For
instance, $x$ might represent the coupling strengths and temperature
of some spin glass model, with the goal being to sample from the
thermal distribution of configurations $y$ for that spin glass.  This
type of computation is particularly useful for simulating the dynamics
of another quantum system.

Unlike Grover's algorithm, Shor's algorithm is usually described as a
sampling computation. The goal is not to directly produce a factor or
list of factors, but rather to produce a distribution over measurement
outcomes.  By sampling from this distribution and doing classical
post-processing it is possible to extract factors of some number $x$.
Of course, as noted in Ref.~\cite{Gershenfeld97a}, it is possible to modify
Shor's algorithm to be a function computation, taking an instance $x$
and producing a list of all the factors of $x$.

The desired result in sampling computation is that the measurement
outcomes $y$ are distributed according to the \emph{ideal}
probabilities $p_x(y)$, for a given problem instance $x$. Suppose,
however, that they are instead distributed according to some nonideal
set of \emph{real} probabilities $q_x(y)$.  How should we compare
these two distributions?  There are two widely-used classical measures
enabling comparison of probability distributions $p$ and $q$.  The
first is the \emph{Kolmogorov} or $l_1$ distance, defined by $D(p,q)
\equiv \sum_y |p(y)-q(y)|/2$.  The second is the \emph{Bhattacharya
  overlap}, defined by $F(p,q) \equiv \sum_y \sqrt{p(y) q(y)}$.  Since
these measures are in fact commutative analogues of the trace distance
and fidelity, respectively, we represent them with the same symbols as
their quantum analogues ($D$ and $F$).  As with the trace distance,
the Kolmogorov distance can be given an appealing interpretation as
the bias in probability when trying to distinguish the distributions
$p$ and $q$.  No similarly simple interpretation for the Bhattacharya
overlap seems to be known, although it is related to the Kolmogorov
distance through inequalities analogous to
Eq.~(\ref{eq:relationship-of-measures}).

The Kolmogorov distance and Bhattacharya overlap, together with the
quantum error measures we have introduced, can be used to relate ideal
and real probability distributions obtained as the result of a quantum
computation.

\textit{Sampling computation in the worst case:} Suppose we attempt to
perform a quantum computation represented by an ideal operation ${\cal
  F}$ that acts on an input $|x\rangle$, where $x$ represents the
instance of the problem to be solved.  The goal is to produce a final
state ${\cal F}(|x\rangle \langle x|)$ which, when measured in the
computational basis, gives rise to an ideal distribution $p_x$.
Instead, we perform the operation ${\cal E}$, giving rise to a
distribution $q_x$ on measurement outcomes.  In
Appendix~\ref{app:sampling:worst} we prove that:
\begin{eqnarray} \label{eq:cssss1}
  \max_x D(q_x,p_x) & \leq & D_{\rm stab}({\cal E},{\cal F})
   \\ \label{eq:cssss2}
  \max_x [1-F(q_x,p_x)] & \leq & C_{\rm stab}({\cal E},{\cal F})^2.
\end{eqnarray}
Just as for function computation, which of these is the better
inequality depends upon the details of the situation under study.

\textit{Sampling computation in the average case:} Given the same
situation as for the worst case, we now assume that problem instances
are chosen uniformly at random.  We will therefore use the Kolmogorov
distance and Bhattacharya overlap between the \emph{joint}
distributions $\{p(x,y)\} \equiv p$ and $\{q(x,y)\} \equiv q$ to
measure how well ${\cal E}$ has approximated ${\cal F}$.  Arguments
analogous to that used in the worst case establish:
\begin{eqnarray} \label{eq:acss1}
  D(q,p) & \leq & D_{\rm pro}({\cal E},{\cal F})
   \\ \label{eq:acss2}
  1-F(q,p) & \leq & C_{\rm pro}({\cal E},{\cal F})^2.
\end{eqnarray}

\section{Summary, recommendations, and conclusion}
\label{sec:conclusion}

We have formulated a list of criteria that must be satisfied by a good measure of error in
quantum information processing.  These criteria provide a
broad framework that can be used to assess candidate error measures,
incorporating both theoretical and experimental desiderata.

We have used this framework to comprehensively survey possible
approaches to the definition of an error measure, rejecting many
\emph{a priori} plausible error measures as they fail to satisfy many
of our criteria.  Although many of these rejected error measures are
of some interest as diagnostic measures, none are suitable for use as
a \emph{primary} measure of the error in a quantum information
processing task.

Four error measures were identified which have particular merit, each
of which satisfies most or all of the criteria we identified.  These
measures are the \emph{J distance} (Jamiolkowski process distance),
the \emph{J fidelity} (Jamiolkowski process fidelity), the
\emph{S distance} (stabilized process distance) and the
\emph{S fidelity} (stabilized process fidelity), denoted $D_{\rm pro},
F_{\rm pro}, D_{\rm stab}$ and $F_{\rm stab}$, respectively.

All four measures either are metrics (in the case of the process
distances) or give rise to a variety of associated metrics (for the
process fidelities). Moreover, all of the metrics can be shown to
satisfy stability and chaining properties which greatly simplify the
analysis of multistage quantum information processing tasks, as
described in the introduction. The main differences arise in the
criteria of easy calculation, measurement and sensible physical
interpretation. We now briefly summarize these remaining properties
for the four measures. Throughout this section, we assume that the
goal in each case is to compare a quantum operation ${\cal E}$ to an
ideal unitary operation $U$; the results vary somewhat when ${\cal E}$
is being compared to an arbitrary process ${\cal F}$.

(i) \textit{J distance:} There is a straightforward formula enabling
  $D_{\rm pro}$ to be calculated directly from the process matrix,
  thus also allowing it to be experimentally determined using quantum
  process tomography.  The \emph{J}~distance can be given an operational
  interpretation as a bound on the average probability of error
  $\overline p_e$ experienced during quantum computation of a
  function, or as a bound on the distance between the real and ideal
  joint distributions of the computer in a sampling computation:
  \begin{eqnarray} \label{eq:jpd1}
  \overline p_e & \leq & \overline p_e^{\rm id} + D_{\rm pro}({\cal E},
  U) \\ \label{eq:jpd2}
  D(q,p) & \leq & D_{\rm pro}({\cal E},U).
  \end{eqnarray}
  In the first expression $\overline p_e^{\rm id}$ is the average
  probability of error in the \emph{ideal} computation, represented by
  $U$.  In the second expression, $D(q,p)$ is the Kolmogorov distance
  between the real joint probability distribution $\{p(x,y)\} \equiv
  p$ on problem instances $x$ and measurement outcomes $y$ and the
  ideal joint distribution $\{q(x,y)\} \equiv q$, for a uniform
  distribution on problem instances.
  
(ii) \textit{J fidelity:} Once again, the \emph{J}~fidelity can be
  calculated directly from the process matrix. However, there is also
  a simpler formula for $F_{\rm pro}$,
  Eq.~(\ref{eq:fidelity-evaluate}), allowing easy calculation and
  measurement, without the need for full process tomography. This is
  much more straightforward than the calculation for the \emph{J}~distance,
  and is likely to simplify the determination of experimental errors.
  As for the \emph{J}~distance, the \emph{J}~fidelity can be given an operational
  interpretation related to average error probabilities:
  \begin{eqnarray} \label{eq:jpf1}
   \overline p_e & \leq & 1-F_{\rm pro}({\cal E},U). \\ \label{eq:jpf2}
  F(q,p) & \geq & F_{\rm pro}({\cal E},U).
  \end{eqnarray}
  In the first expression we are now restricted to ideal computations
  $U$ which succeed perfectly, i.e., $\overline p_e^{\rm id} = 0$.  In
  the second expression, $F(q,p)$ is the Bhattacharya overlap between
  the real and ideal joint probability distributions, $p$ and $q$,
  again for a uniform distribution on problem instances.
  
(iii) \textit{S distance:} There is no known elementary formula for
  $D_{\rm stab}$, but we have proved that calculating the \emph{S}~distance
  is equivalent to a convex optimization problem, which can be
  efficiently solved numerically, given knowledge of the process.
  This, in turn, enables $D_{\rm stab}$ to be measured experimentally,
  by performing full quantum process tomography.  The \emph{S}~distance can
  be simply interpreted as a bound on the worst-case error probability
  $p_e$ for a function computation, and as a bound on the maximum
  distance between the real and ideal output distributions of a
  sampling computation:
  \begin{eqnarray} \label{eq:spd1}
   p_e & \leq & p_e^{\rm id}+ D_{\rm stab}({\cal E},
        U). \\ \label{eq:spd2}
  \max_x D(q_x,p_x) & \leq & D_{\rm stab}({\cal E},U).
  \end{eqnarray}
  In the first expression $p_e^{\rm id}$ is the worst-case error
  probability in the ideal computation, $U$.  In the second expression
  $D(q_x,p_x)$ is the Kolmogorov distance between the real and ideal
  output probability distributions $\{q_x(y)\} \equiv q_x$ and $p_x$,
  and we take the worst case over all problem instances $x$.
  
(iv) \textit{S fidelity:} Once again, no elementary formula for the
  \emph{S}~fidelity is known, but we have proved that the determination of
  $F_{\rm stab}$ can be formulated as a convex optimization problem,
  and thus $F_{\rm stab}$ can be efficiently determined numerically.
  As a result, $F_{\rm stab}$ can again be determined experimentally,
  using process tomography.  As with the \emph{S}~distance, $F_{\rm pro}$ has
  an operational interpretation related to worst-case error
  probabilities:
  \begin{eqnarray} \label{eq:spf1}
  p_e & \leq & \left( \sqrt{p_e^{\rm id}} + C_{\rm stab}({\cal E},
        U) \right)^2. \\ \label{eq:spf2}
  \min_x F(q_x,p_x) & \geq & F_{\rm stab}({\cal E},U).
  \end{eqnarray}
  The notation here is the same as above, with the definition
  $C_{\rm stab}({\cal E},U) \equiv \sqrt{1-F_{\rm stab}({\cal E},U)}$.

%
%
Which of these four error measures is the best?  Our recommendation is
necessarily tentative, for we do not yet have a complete understanding
of the properties of these measures.  In particular, the discovery of
simpler formulae for calculating the measures or simpler procedures
for measuring them experimentally remain possibilities which could
make it necessary to reconsider their relative merits.

%
%
The fact that they all four measures obey the stability and chaining
criteria means that in all cases it is only necessary to characterize
the component processes in order to bound the total error in a complex
quantum information processing task. This makes conceivable the idea
of using these measures for assessing processes in large-scale
systems.

%
%
One important difference between the measures is that the \emph{S}~distance
and \emph{S}~fidelity bound worst-case error probabilities, as compared to
the average-case error probabilities for which the \emph{J}~distance and
\emph{J}~fidelity provide bounds. This would seem to be a significant
advantage for the \emph{S}~distance and \emph{S}~fidelity, since worst-case errors
are usually of more interest than the average case. On the other hand,
given the linear nature of quantum mechanics, it seems likely that in
low dimensions relatively tight ways may be found to use the average
errors to bound the worst-case errors.

%
%
The measure which is simplest to calculate is the \emph{J}~fidelity, which
has a simple formula, and is relatively easy to determine
experimentally compared with the other measures.  Unfortunately, this
measure has the weakest operational interpretation of the four.  As
well as being only related to the average-case probability of error,
our expression Eq.~(\ref{eq:jpf1}) does not hold true for function
computations where the ideal case suffers an intrinsic error. For this
reason we believe that the \emph{J}~fidelity is of particular interest for
early, proof-of-principle experimental demonstrations, but that other
measures with more desirable properties will eventually supersede it.

%
%
The \emph{J}~distance has different strengths and weaknesses than the
\emph{J}~fidelity. On the one hand, it does allow the analysis of function
computations with intrinsic errors in the ideal case. However, it
requires a full process tomography to be determined experimentally, it
is not as easy to calculate, and is still only related to average
errors.

%
%
The \emph{S}~distance and \emph{S}~fidelity have the most attractive operational
interpretations, since they relate to worst-case error probabilities.
Unfortunately, they are also more difficult to determine
experimentally than the \emph{J}~fidelity, requiring full process tomography,
and no elementary formula for either is known.  However, they are easy
to calculate numerically, and although full process tomography is a
time-consuming task, it is becoming a standard technique in quantum
information experiments. 

On the basis of their compelling operational interpretations, and
other attractive theoretical and experimental properties, we believe
that the \emph{S}~distance and \emph{S}~fidelity are the two best error
  measures, and should be used as the basis for comparison of real
  quantum information processing experiments to the theoretical
  ideal.  

%
%
Is it possible to make a definite recommendation as regards which of
these two measures to use? At the moment, we know of no convincing
argument to choose one over the other. For instance, it is
straightforward to find examples of different processes where either
the \emph{S}~distance or the \emph{S}~fidelity give the better bound in
Eqs.~(\ref{eq:spd1}) and~(\ref{eq:spf1}).  Further work on the
relative merits of these measures is required before a definitive
choice can be made.

As a consequence, at the present time we believe that \emph{both}
measures should be reported in experiments.  Note that determining two
measures rather than one imposes little additional burden on
experimentalists, since determining either measure requires (at
present) process tomography to be performed, and once process
tomography has been performed it is straightforward to numerically
calculate both measures.

%
%
Much work remains to be done.  Tasks of obvious importance include:
(a) obtaining closed-form formulae and simple experimental measurement
procedures for the \emph{S}~distance and \emph{S}~fidelity; (b) finding procedures
which can be used to calculate experimental error bars for the
\emph{S}~distance and \emph{S}~fidelity; (c) expressing the threshold condition for
fault-tolerant quantum computation and communication using the error
measures we have identified; and (d) extending our work so that it
applies to quantum operations which are not trace-preserving, such as
arise naturally in certain optical proposals for quantum
computation~\cite{Knill01a,Nielsen04b}, where measurements and
post-selection are critical elements.

Broadening the scope, it would also be useful to develop additional
diagnostic measures, which could be used experimentally to understand
and improve specific aspects of a process's operation, while not being
suitable as general-purpose measures of how well a process has been
performed.  An example of such a measure is the \emph{process purity},
$\mbox{tr}(\rho_{\cal E}^2)$, which can be regarded as a measure of
the extent to which a quantum operation ${\cal E}$ maintains the
purity of the quantum state.  Although this measure is easily seen to
be deficient in terms of the criteria developed in the introduction,
and thus is not suitable as a general-purpose measure, it may be
useful as a diagnostic measure that provides information about one
specific aspect of ${\cal E}$'s performance.

\acknowledgments

AG acknowledges support from the New Zealand Foundation for Research,
Science and Technology under grant UQSL0001.  MAN thanks Carl Caves,
who has repeatedly emphasized the significance of obtaining suitable
criteria for quantum information processing.

\appendix
\section{Worst case proofs}
\subsection{Proof of worst-case stabilization}
\label{app:worst-case-stabilization}

Let $\mathcal{E}_Q$ and $\mathcal{F}_Q$ be trace-preserving quantum
operations acting on a $d$-dimensional system $Q$.  We will show,
following Ref.~\cite{Aharonov98a}, that $\Delta_{\rm stab}({\cal E}_Q,{\cal
  F}_Q)$ is stable under the addition of an arbitrary $d'$-dimensional
system $Q'$, i.e, $\Delta_{\rm stab}({\cal E}_Q,{\cal F}_Q) =
\Delta_{\rm stab}({\cal I}_{Q'} \otimes {\cal E}_{Q},{\cal I}_{Q'}
\otimes {\cal F}_Q)$

To see this, recall the definition of $\Delta_{\rm stab}({\cal
  E}_Q,{\cal F}_Q)$.  We introduce a fictitious $d$-dimensional
ancillary system $A$, acted upon by the identity operation
$\mathcal{I}_A$.  Then by definition $\Delta_{\rm stab}({\cal
  E}_Q,{\cal F}_Q) \equiv \Delta_{\max}({\cal I}_A \otimes {\cal
  E}_Q,{\cal I}_A \otimes {\cal F}_Q)$.

%
%
By definition of $\Delta_{\rm stab}$ we see that $\Delta_{\rm
  stab}({\cal I}_{Q'} \otimes {\cal E}_{Q},{\cal I}_{Q'} \otimes {\cal
  F}_Q)$ is equal to $\Delta_{\max}({\cal I}_{B}\otimes{\cal I}_{Q'}
\otimes {\cal E}_Q,{\cal I}_{B}\otimes{\cal I}_{Q'} \otimes {\cal
  F}_Q)$, where ${\cal I}_{B}$ acts as the identity on a $d\times
d'$-dimensional ancilla $B$. Thus, to prove stability it suffices to
show that the quantity $\Delta_{\max}({\cal I}_S \otimes {\cal
  E}_Q,{\cal I}_S \otimes {\cal F}_Q)$ is independent of the dimension
of the system $S$ that ${\cal I}_S$ acts on, provided $S$ is at least
$d$-dimensional.

%
%
To see this independence, let $\psi$ be a state achieving the maximum
in $\Delta_{\max}({\cal I}_S \otimes {\cal E}_Q,{\cal I}_S \otimes
{\cal F}_Q)$, with a Schmidt decomposition $\psi = \sum_j \psi_j
|e_j\rangle |f_j\rangle$, where $|e_j\rangle$ are orthonormal states
of $S$, and $|f_j\rangle$ is an orthonormal basis set for $Q$.  Since $Q$ is
$d$-dimensional, the state $\psi$ has at most $d$ Schmidt
coefficients, and so we can restrict our attention to that
$d$-dimensional subspace of $S$ spanned by the states $|e_j\rangle$
with nonzero Schmidt coefficients.  We see that the maximum can be
obtained working only in this subspace, concluding the proof.

\subsection{Proof of convex optimization property for $F_{\rm stab}$}
\label{app:proof-conv-opt}
%
%
Our goal is to show that the problem of computing $F_{\rm stab}$ can
be reduced to the minimization of a convex function defined on a
convex set.  To show this we introduce a new function, denoted
$F(\rho_Q, {\cal E}_Q,{\cal F}_Q)$, where subscripts indicate the
system on which the variable is defined. The value of $F(\rho_Q, {\cal
  E}_Q,{\cal F}_Q)$ is defined to be the state fidelity $F(({\cal I}_A
\otimes {\cal E}_Q)(\psi),({\cal I}_A \otimes {\cal F}_Q)(\psi))$,
where $A$ is an ancilla of at least the same dimension as $Q$, and
$\psi$ is any purification of $\rho_Q$ to $AQ$.  It is easily verified
that this definition is independent of which purification $\psi$ of
$\rho_Q$ is used.

%
%
From this definition, it can be seen that the problem of computing
$F_{\rm stab}({\cal E}_Q,{\cal F}_Q)$ is equivalent to minimizing
$F(\rho_Q,{\cal E}_Q,{\cal F}_Q)$ over all density matrices $\rho_Q$
of system $Q$.  Therefore, to prove that finding $F_{\rm stab}$ is a
convex optimization problem, we simply need to show that
$F(\rho_Q,{\cal E}_Q,{\cal F}_Q)$ is a convex function of $\rho_Q$,
which takes values in a convex set.

%
%
To do this, let $p_j$ be probabilities, and let $\rho_Q^j$ be
corresponding states of the system $Q$, with purifications $\psi_j$ to
a system $AQ$.  It is helpful to introduce another ancillary system
$A'$ with an orthonormal basis $|j\rangle$ in one-to-one
correspondence with the index on the states $\rho_Q^j$, and we define
a state $|\psi\rangle \equiv \sum_j \sqrt{p_j} |j\rangle
|\psi_j\rangle$ of the joint system $A'AQ$.  By observing that
$|\psi\rangle$ is a purification of $\sum_j p_j \rho_Q^j$, we see that
\begin{eqnarray}
\nonumber
&&  F\left(\sum_j p_j \rho_Q^j,{\cal E}_Q,{\cal F}_Q \right) \\
&& \quad =
  F(({\cal I}_{A'A} \otimes {\cal E}_Q)(\psi),({\cal I}_{A'A} \otimes
  {\cal F}_Q)(\psi)).
\end{eqnarray}
We then apply the monotonicity of the fidelity (c.f.
Sec.~\ref{sec:distance-states}) under decoherence in the $|j\rangle$
basis, giving
\begin{eqnarray}
  & & F\left( \sum_j p_j \rho_Q^j,{\cal E}_Q,{\cal F}_Q \right)  \leq 
   \nonumber \\
  & & \hphantom{abc} F\left( \sum_j p_j |j\rangle \langle j| \otimes 
  ({\cal I}_{A} \otimes {\cal E}_Q)(\psi_j), \right. \nonumber \\
  & & \hphantom{abc F \left( \right. } \left. \sum_j p_j |j\rangle \langle j|
  \otimes ({\cal I}_{A} \otimes
  {\cal F}_Q)(\psi_j)\right). 
\end{eqnarray}
Finally, applying some elementary algebra to simplify the right-hand
side, we obtain
\begin{eqnarray}
  F\left(\sum_j p_j \rho_Q^j,{\cal E}_Q,{\cal F}_Q\right) & \leq &
  \sum_j p_j F( \rho^j_Q, {\cal E}_Q, {\cal F}_Q), \nonumber \\
  & & 
\end{eqnarray}
which implies that $F(\rho_Q,{\cal E}_Q,{\cal F}_Q)$ is convex in
$\rho_Q$, as desired.

%
%
A similar construction shows that the computation of $D_{\rm stab}$ is
equivalent to the maximization of a concave function over a convex
set, and thus is also a convex optimization problem, with concomitant
numerical benefits.  The construction is sufficiently similar that we
omit the details.

\section{Application to quantum computing}
\label{app:application}

\subsection{Function computation in the worst case}
\label{app:function:worst}

Suppose ${\cal E}$ and ${\cal F}$ are real and ideal quantum
operations, respectively, that act on an input $|x\rangle$, where $x$
represents a problem instance. ${\cal E}$ succeeds in computing the
desired function $f(x)$ with an error probability of at most $p_e$,
whereas ${\cal F}$ succeeds with an (ideal) error probability of at
most $p_e^{\rm id}$.

We wish to show:
\begin{eqnarray} \label{app:eq:fcsss1}
  p_e & \leq & p_e^{\rm id} + D_{\rm stab}({\cal E},{\cal F}) \\
\label{app:eq:fcsss2}
  p_e & \leq & \left( \sqrt{p_e^{\rm id}} + C_{\rm stab}({\cal E},
        {\cal F}) \right)^2.
\end{eqnarray}

%
%
To prove the first inequality,~(\ref{app:eq:fcsss1}), we introduce a
quantum operation ${\cal M}$ representing the process of measurement,
${\cal M}(\rho) = \sum_y |y\rangle \langle y| \rho |y\rangle \langle
y|$, where the sum is over all possible measurement outcomes $y$.  Now
observe that
\begin{eqnarray}
  p_e & = & 
            D(({\cal M} \circ {\cal E})(|x\rangle \langle x|),
            |f(x)\rangle \langle f(x)|) \\
 & \leq & D(({\cal M} \circ {\cal E})(|x\rangle \langle x|),
          ({\cal M} \circ {\cal F})(|x\rangle \langle x|)) \nonumber \\
 & & 
        +D(({\cal M} \circ {\cal F})(|x\rangle \langle x|)),
            |f(x)\rangle \langle f(x)|) \\
 & \leq & D({\cal E}(|x\rangle \langle x|),
          {\cal F}(|x\rangle \langle x|))
        +p_e^{\rm id},
\end{eqnarray}
where we used simple algebra in the first line, the triangle
inequality in the second line, and contractivity of trace distance and
some simple algebra in the third line.  The desired result,
Eq.~(\ref{app:eq:fcsss1}), now follows from the definition of $D_{\rm stab}$.

%
%
To prove the second inequality, Eq.~(\ref{app:eq:fcsss2}), note that
\begin{eqnarray}
  p_e & = & 1 - F({\cal E}(|x\rangle \langle x|),|f(x) \rangle \langle f(x)|) \\
& = & C({\cal E}(|x\rangle \langle x|),|f(x) \rangle \langle f(x)|)^2 \\
 & \leq & [ C({\cal E}(|x\rangle \langle x|),
 {\cal F}(|x\rangle \langle x|)) \nonumber \\
& & + \; C({\cal F}(|x\rangle\langle x|),|f(x)\rangle \langle f(x)| ]^2,
\end{eqnarray}
where the first line follows from the definition of $p_e$ and the
state fidelity, the second line follows from the definition of the
metric $C(\cdot,\cdot)$, and the third line follows from the triangle
inequality for $C(\cdot,\cdot)$.  The proof of Eq.~(\ref{app:eq:fcsss2})
is completed by noting that $C({\cal E}(|x\rangle \langle x|),{\cal
  F}(|x\rangle \langle x|)) \leq C_{\rm stab}({\cal E},{\cal F})$ and
$C({\cal F}(|x\rangle \langle x|),|f(x) \rangle \langle f(x)|) \leq \sqrt{p_e^{\rm id}}$.

\subsection{Function computation in the average case}
\label{app:function:average}

As in the worst case, ${\cal E}$ and ${\cal F}$ are real and ideal
quantum operations that act on an input $|x\rangle$ to compute a
desired function $f(x)$. ${\cal E}$ succeeds with an average error
probability $\overline{p}_e$, whereas ${\cal F}$ succeeds with an
average error probability $\overline{p}_e^{\rm id}$.

The first steps in the proof of
Eq.~(\ref{eq:fcacss1}) are directly analogous to the proof of
Eq.~(\ref{eq:fcsss1}), resulting in the inequality
\begin{eqnarray}
  \overline p_e & \leq & \overline p_e^{\rm id} + \frac{1}{d}
  \sum_x D({\cal E}(|x\rangle \langle x|),{\cal F}(|x\rangle \langle x|)),
\end{eqnarray}
where $d$ is the total number of possible inputs $x$.  Recall that
\begin{eqnarray}
  D_{\rm pro}({\cal E},{\cal F}) & = & D(({\cal I} \otimes {\cal E})
  (\Phi),({\cal I} \otimes {\cal F})
  (\Phi), 
\end{eqnarray}  
where ${\cal I}$ acts on an ancilla which is a copy of the system
${\cal E}$ and ${\cal F}$ act on, and $|\Phi\rangle =\sum_x |x\rangle
|x\rangle/ \sqrt{d}$ is a maximally entangled state of the two
systems.  Now let ${\cal M}$ be a quantum operation representing
measurement on the ancilla system, defined similarly to the definition
of ${\cal M}$ just above.  By contractivity of the trace distance,
\begin{eqnarray}
  D_{\rm pro}({\cal E},{\cal F}) \geq D(({\cal M} \otimes {\cal E})
  (\Phi),({\cal M} \otimes {\cal F})
  (\Phi)). 
\end{eqnarray}  
Elementary algebra gives
\begin{eqnarray}
  & & D(({\cal M} \otimes {\cal E})
  (\Phi),({\cal M} \otimes {\cal F})
  (\Phi)) \nonumber \\
 & = & \frac{1}{d} \sum_x 
  D({\cal E}(|x\rangle \langle x|),{\cal F}(|x\rangle \langle x|)).
\end{eqnarray}  
Combining these results, we obtain Eq.~(\ref{eq:fcacss1}).

%
%
As already remarked we have not found a natural average-case analogue
of Eq.~(\ref{eq:fcsss2}).  However, if $\overline p_e^{\rm id} = 0$,
i.e., our computation succeeds with probability one, then it is
possible to prove an average-case analogue.  The result is
\begin{eqnarray}   \label{eq:fcacss3}
  \overline p_e & \leq & C_{\rm pro}({\cal E},
        {\cal F})^2 = 1-F({\cal E},{\cal F}).
\end{eqnarray}
The proof uses very similar techniques to those used to establish
Eqs.~(\ref{eq:fcacss1}) and~(\ref{eq:fcsss2}), and is therefore
omitted.

\subsection{Sampling computation in the worst case}
\label{app:sampling:worst}

The quantum operation ${\cal E}$ is an imperfect attempt to reproduce
the statistics of the ideal operation ${\cal F}$ which acts on an
input $|x\rangle$. Measured in the computational basis, ${\cal F}$
gives rise to a distribution $\{p_x(y)\} \equiv p_x$, whereas ${\cal
  E}$ gives a distribution $\{q_x(y)\} \equiv q_x$.

The inequalities Eqs.~(\ref{eq:cssss1}) and~(\ref{eq:cssss2}) that we
want to prove may be stated as follows:
\begin{eqnarray} \label{eq:cssss1a}
  \max_x D(q_x,p_x) & \leq & D_{\rm stab}({\cal E},{\cal F})
   \\ \label{eq:cssss2a}
  \min_x F(q_x,p_x) & \geq & F_{\rm stab}({\cal E},{\cal F}).
\end{eqnarray}
To prove the first inequality,~(\ref{eq:cssss1a}), let ${\cal M}$ again
be a quantum operation representing measurement in the computational
basis.  Note that for all $x$
\begin{eqnarray}
  D(q_x,p_x) & = & D(({\cal M} \circ {\cal E})(|x\rangle
  \langle x|),({\cal M} \circ {\cal F})(|x\rangle \langle x|)) \nonumber \\
  & & \\
  & \leq & D({\cal E}(|x\rangle \langle x|),{\cal F}(|x\rangle \langle x|)) \\
  & \leq & D_{\rm stab}({\cal E},{\cal F}),
\end{eqnarray}
where we used simple algebra in the first line, contractivity in the
second line, and the definition of $D_{\rm stab}$ in the third line.
An analogous argument can be used to establish the second
inequality,~(\ref{eq:cssss2a}).



\begin{thebibliography}{55}
\bibitem[{foo({\natexlab{a}})}]{footnote1}
\bibinfo{note}{Our use of this term was suggested by a talk of
  Caves~\cite{Caves01a}; see also Ref.~\cite{Caves04a}, which discusses possible
  gold standards for quantum teleportation.}
\bibitem[{foo({\natexlab{d}})}]{footnote4}
\bibinfo{note}{The operator inequality $A \leq B$ means that $B-A$ is a
  positive matrix. Thus $A \leq I$ means that $A$ is a Hermitian matrix whose
  eigenvalues are at most $1$.}
\bibitem[{foo({\natexlab{e}})}]{footnote5}
\bibinfo{note}{Post-selection is usually associated with some specific
  measurement outcome being obtained.}
\bibitem[{foo({\natexlab{f}})}]{footnote6}
\bibinfo{note}{An $n$-qubit example of an orthogonal operator basis is the set
  of all possible tensor products formed from the Pauli matrices and the
  identity. These operators may be normalized by multiplying by an appropriate
  constant, giving an orthonormal operator basis.}
\bibitem[{foo({\natexlab{h}})}]{footnote8}
\bibinfo{note}{Note, however, that not all density matrices may be realized as
  states $\rho_{\cal E}$. It turns out that the class ${\cal E}$ of quantum
  states that may arise in this way from a trace-preserving operation is simply
  all those $\rho_{\cal E}$ whose reduced density matrix on the copy of the
  original system is the completely mixed state, $I/d$~\cite{Horodecki99c}.}
\bibitem[{foo({\natexlab{i}})}]{footnote9}
\bibinfo{note}{A review of the properties of the trace distance and fidelity
  may be found in Chap.~9 of Ref.~\cite{mikenike}.}
\bibitem[{foo({\natexlab{k}})}]{footnote11}
\bibinfo{note}{It is noteworthy that using the square root of the definition in
  Eq.~(\ref{eq:fidelity}) results in ``better'' values for the fidelity, and
  this may account for some of the attractiveness of that quantity. Needless to
  say, this is not a compelling reason for using it.}
\bibitem[{foo({\natexlab{l}})}]{footnote12}
\bibinfo{note}{Strictly, this statement is only true in projective state space,
  in which states which are the same up an overall phase are regarded as
  identical.}
\bibitem[{foo({\natexlab{m}})}]{footnote13}
\bibinfo{note}{To prove the triangle inequality for $C(\rho,\sigma)$, we note
  that by Uhlmann's theorem (see Ref.~\cite{Uhlmann76a} or Theorem~9.4 on page~410
  of Ref.~\cite{mikenike}) given any three states $\rho,\sigma$ and $\tau$ there
  exist corresponding purifications $|r\rangle, |s\rangle$ and $|t\rangle$ such
  that $F(\rho,\sigma) = |\langle r|s\rangle|^2, F(\rho,\tau) = |\langle
  r|t\rangle |^2$, and $F(\sigma,\tau) = |\langle s|t\rangle|^2$. Let
  $\theta_{rs}$ be an angle in the range $0$ to $\pi/2$ so that
  $\cos(\theta_{rs}) = |\langle r|s\rangle|$. Define $\theta_{rt}$ and
  $\theta_{st}$ similarly. With these definitions it is easy to check that
  $C(\rho,\sigma) = \sin(\theta_{rs})$, with similar relations holding for the
  other pairs of density matrices, and so the triangle inequality reduces to
  proving $\sin(\theta_{rt}) \leq \sin(\theta_{rs})+\sin(\theta_{st})$. But
  elementary geometry implies that $\theta_{rt} \leq \theta_{rs} +
  \theta_{st}$, and simple algebra can be used to deduce the triangle
  inequality from that.}
\bibitem[{foo({\natexlab{n}})}]{footnote14}
\bibinfo{note}{This observation seems to be folklore. We do not know where it
  originated, but a proof is provided in Ref.~\cite{02nielsen249}.}
\bibitem[{foo({\natexlab{o}})}]{footnote15}
\bibinfo{note}{Note that the $\sigma_l$ defined in this way are not, in
  general, Hermitian, and thus may not be observables. However, we can easily
  split them into the form $\sigma_l = A_l + i B_l$, where $A_l$ and $B_l$ are
  Hermitian, and thus measuring $\mbox{tr}(\sigma_l \rho)$ can be achieved
  simply by measuring two separate observables.}
\bibitem[{foo({\natexlab{p}})}]{footnote16}
\bibinfo{note}{Obviously, taking the maximum over pure states will be
  numerically easier as it will involves only $2d-2$ real parameters for a
  $d$-dimensional state space, as opposed to $d^2-1$ for a normalized mixed
  states.}
\bibitem[{foo({\natexlab{r}})}]{footnote18}
\bibinfo{note}{One can consider an alternate model in which the starting state
  of the computer is some standard state, $|s\rangle$, and the operation to be
  performed ${\cal F}_x$ depends on the problem instance. Exactly analogous
  results to those we will describe here can be proved in this alternate
  context.}
\end{thebibliography}

\end{document}